\documentclass[english]{article}
\usepackage[T1]{fontenc}
\usepackage[latin9]{inputenc}
\usepackage{float}
\usepackage{amssymb}
\usepackage{graphicx}
\usepackage{esint}

\makeatletter
\newcommand{\lyxaddress}[1]{
\par {\raggedright #1
\vspace{1.4em}
\noindent\par}
}

\@ifundefined{showcaptionsetup}{}{%
 \PassOptionsToPackage{caption=false}{subfig}}
\usepackage{subfig}
\makeatother

\usepackage{babel}
\begin{document}

\title{Collapse and revival for a slightly anharmonic Hamiltonian}

\author{Alexandra Bakman, Hagar Veksler and Shmuel Fishman}

\maketitle

\lyxaddress{Technion - Israel Institute of Technology, Technion City, Haifa 3200003,
Israel}
\begin{abstract}
The effect of quantum collapse and revival is a fascinating interference
phenomenon. In this paper the phenomenon is demonstrated analytically
and numerically for a simple system, a slightly anharmonic Hamiltonian.
The initial wave-function is a displaced ground state of a harmonic
oscillator. Possible experimental realizations for cold atoms are
discussed in detail.
\end{abstract}

\section{Introduction\label{sec:Introduction}}

Collapse and revival phenomena are fascinating and are encountered
in a variety of physical situations, that were explored experimentally
and theoretically. In the present paper a simple example that can
be analyzed analytically, and may be realized experimentally, is presented. 

The first collapse and revival phenomenon that was observed and explained
theoretically is the Talbot effect \cite{Talbot_original,Rayleigh_Talbot},
where the amplitude of an optical signal collapses and then revives
partially and completely. A quantum phenomenon of this type is the
quantum carpet \cite{Quantum_Carpets_Light_Berry,Quantum_Carpets_Made_Simple,Quantum_Carpet_Wigner}.
Some revivals discussed here are of different nature than the ones
found for the Talbot effect and the quantum carpet. Collapses and
revivals, as well as fractional revivals and superrevival structures
were observed for wave packets in Rydberg atoms \cite{ORIGINAL_Yeazell_Stroud_Collapse_Revival,(1)Rydberg_Atoms_Collapse_Revival,Rydberg_Revivals,Rydberg_Superrevivals,Fractional_Revivals_Averbukh_Perelman,ORIGINAL_Averbukh,ORIGINAL_Averbukh_Perelman,Walter_Collapse_Revival}.
Also, this phenomenon was observed for interacting bosons \cite{Bloch_Collapse_Revival,Phillips_Revivals}
, and model systems \cite{Jaynes_Cummings,(0)_Periodic_Collapse_Revival,Bose-Hubbard_Numeric_1,Bose-Hubbard_2,Fractional_Jaynes_Cumming}.
Similar phenomenon was found for chaotic systems \cite{(2)Chaotic_Collapse_Revival}.
For the two site Bose Hubbard model, the difference in populations
of the two sites exhibits collapses and revivals \cite{Hagar_Collapse_Revival}. 

A simple model where the phenomenon of collapses and revivals is found
is for noninteracting bosons in a weakly anharmonic trap \cite{Mark_Revivals}.
In this specific situation the particles are prepared in the ground
state of the trap and then the potential is instantaneously shifted
by some distance. In a harmonic trap, the wave packet oscillates with
the frequency of the trap and so do the various observables, for example,
the position $x$ and the momentum $p.$ As a result of the anharmonicity,
these oscillations are superimposed by an envelope exhibiting collapses
and revivals. Some of our numerical results were presented in \cite{Mark_Revivals}
that focused on a different issue, namely echoes resulting of the
interplay between two displacements. In the present work, analytic
formulas for the evolution of the observables are derived. 

One should remember that within the model we explore the evolution
is coherent and information is not lost even during the collapse.
This coherence enables the revivals. 

The model presented here is simple and the evolution of the observables
is described in a straightforward manner. The simplicity is of great
value if used to explore more complex situations. For example, effects
of interparticle interactions will result in deviations from our predictions.
These can be detected in experiments. The significance of these effects
can be tuned by the particle density. 

The analytical method we use in the present paper is the semiclassical
approximation assuming that the energy level spacing is much smaller
than the energy. 

When we expand a symmetric potential around its minimum, the leading
order is harmonic and the first correction is a term of the form $\beta x^{4}.$
Therefore, we study a model described by the Hamiltonian

\begin{equation}
H^{'}=\frac{p^{\prime2}}{2m^{\prime}}+\frac{1}{2}m^{\prime}\omega_{0}^{\prime2}x^{\prime2}+\frac{\beta^{\prime}}{4}x^{\prime4},\label{eq:Unscaled_Hamiltonian}
\end{equation}
In dimensionless units

\begin{eqnarray}
x & = & x^{\prime}/\sqrt{\hbar/m^{\prime}\omega_{0}^{\prime}}\label{eq:x_rescaled}\\
t & = & \omega_{0}^{\prime}t^{\prime}\label{eq:t_rescaled}\\
H & = & \frac{H^{\prime}}{\hbar\omega_{0}^{\prime}}\label{eq:H_rescaled}\\
p & = & p^{\prime}\sqrt{\frac{1}{\hbar m^{\prime}\omega_{0}^{\prime}}},\label{eq:p_rescaled}
\end{eqnarray}

\noindent where prime denotes the corresponding values in physical
units, and the Hamiltonian takes the form

\begin{equation}
H=\frac{p^{2}}{2}+\frac{1}{2}x^{2}+\frac{1}{4}\beta x^{4},\label{eq:Rescaled_Hamiltonian}
\end{equation}

\noindent where 
\begin{equation}
\beta=\beta^{\prime}\left(\frac{\hbar}{m^{\prime2}\omega_{0}^{\prime3}}\right).\label{eq:beta_rescaled-1}
\end{equation}
 In the present work we assume $\beta\ll1.$ The Schrödinger equation
is

\begin{equation}
i\frac{\partial}{\partial t}\psi=H\psi.\label{eq:Schr=0000F6dinger_equation}
\end{equation}

The model (\ref{eq:Rescaled_Hamiltonian}) is an idealized model that
will be shown to exhibit collapse and revival behavior. This is a
very general phenomenon, and therefore it is relevant for a large
variety of models. More generally, let us expand eigenenergies around
a level $\bar{n}$ \cite{Quantum_Revivals_Robinett,Revival_Structure_Bluhm_Kostelecky}
in the form 

\begin{equation}
E_{n}\simeq E_{\bar{n}}+E_{\bar{n}}^{\prime}\delta+\frac{1}{2}E_{\bar{n}}^{\prime\prime}\delta^{2}+\frac{1}{6}E_{\bar{n}}^{\prime\prime\prime}\delta^{3}\label{eq:energy_expansion}
\end{equation}

\noindent where $\delta$ is defined by $\delta=n-\bar{n}$ , $\bar{n}$
is the state with maximal probability for a displaced ground state
of a harmonic oscillator, $E_{\bar{n}}^{\prime},$ $E_{\bar{n}}^{\prime\prime}$
and $E_{\bar{n}}^{\prime\prime\prime}$ are derivatives of the energy
with respect to the quantum number $n,$ calculated at $\bar{n}.$
We consider a situation where the derivatives decrease rapidly with
the order, so that

\begin{equation}
E_{\bar{n}}^{\prime}\gg E_{\bar{n}}^{\prime\prime}\gg E_{\bar{n}}^{\prime\prime\prime}.\label{eq:energy_hierarchy}
\end{equation}
 This is the case for high energy levels $n$ if the level is to a
good approximation a power of the quantum number \cite{Rydberg_Revivals}.
From condition (\ref{eq:energy_hierarchy}) follows the separation
of time scales in the system dynamics. 

In Section \ref{sec:Collapse-and-Revival} the dynamics of observables
is calculated for the simple model (\ref{eq:Rescaled_Hamiltonian})
while in Section \ref{sec:Experimental} possible experimental realizations
in the field of cold atoms are presented. The results are summarized
and discussed in Section \ref{sec:Discussion}.

\section{Collapse and Revival of the expectation values of position and momentum\label{sec:Collapse-and-Revival}}

In this section we study the expectation values of the position and
momentum operators for an initial Gaussian wavepacket displaced by
some distance $d$ from the minimum of the potential. For small $\beta$
the system exhibits collapses and revivals \cite{Mark_Revivals}.
This phenomenon is explained in terms of the semiclassical approximation
using a method similar to the one used in \cite{Hagar_Collapse_Revival}.
The phenomenon was found numerically and semianalytically for the
system (\ref{eq:Unscaled_Hamiltonian}) in \cite{Mark_Revivals}.
Since this phenomenon is expected to take place for high energy levels,
where the energy is much larger than the level spacing, we use the
semiclassical approximation in the leading order. 

It is important to remember that second order in $\beta$ of the semiclassical
approximation is more accurate than second order in perturbation theory
for high energy levels. (see detailed discussion in \cite{Hagar_Collapse_Revival})
Using the semiclassical spectrum the evolution of $\left\langle \hat{x}\left(t\right)\right\rangle $
and $\left\langle \hat{p}\left(t\right)\right\rangle $ is calculated.

\subsection{Energy spectrum calculation using the WKB (Wentzel, Kramers and Brillouin)
approximation\label{sub:Energy-spectrum-calculation}}

In order to use the WKB approximation, we calculate the action integral
as

\begin{eqnarray}
I & = & \frac{1}{\pi}\int_{-\widetilde{a}}^{\widetilde{a}}pdx\label{eq:Action_Integral}\\
 & = & \frac{1}{\pi}\int_{-\widetilde{a}}^{\widetilde{a}}\sqrt{2\left(E-\frac{1}{2}x^{2}-\frac{\beta}{4}x^{4}\right)}dx\nonumber 
\end{eqnarray}

\noindent where $\pm\widetilde{a}$ are the turning points of the
path related to the energy by

\begin{equation}
E=\frac{1}{2}\widetilde{a}^{2}+\frac{\beta}{4}\widetilde{a}^{4}.\label{eq:Energy-1}
\end{equation}

\noindent The integral is calculated to the second order in $\beta$
and is found to be 

\begin{eqnarray}
I & = & E-\frac{3}{8}\beta E^{2}+\frac{35}{64}\beta^{2}E^{3}.\label{Action vs. Energy}
\end{eqnarray}

\noindent This is the value of the action for a fixed value of energy
and $\beta.$ Solving for the energy as function of the action to
the second order in $\beta$ results in

\begin{equation}
E=I+\frac{3}{8}\beta I^{2}-\frac{17}{64}\beta^{2}I^{3}.\label{eq:Energy Spectrum Action}
\end{equation}
The action is quantized as

\begin{equation}
I_{n}=n+\frac{1}{2}.\label{eq:Action_Quantization}
\end{equation}

\noindent Substituting (\ref{eq:Action_Quantization}) in (\ref{eq:Energy Spectrum Action}),
the energy spectrum of the Hamiltonian to second power of $\beta$
yields

\begin{eqnarray}
E_{n} & = & \left(n+\frac{1}{2}\right)+\frac{3\beta}{8}\left(n^{2}+n+\frac{1}{4}\right)\label{eq:Energy_spectrum_full}\\
 & - & \beta^{2}\left(\frac{17}{64}n^{3}+\frac{51}{128}n^{2}+\frac{51}{256}n+\frac{17}{512}\right).\nonumber 
\end{eqnarray}

A slightly different spectrum is obtained by using quantum perturbation
theory. Comparison to the exact result obtained by numerical diagonalization
of the Hamiltonian (\ref{eq:Rescaled_Hamiltonian}) shows that the
WKB approximation gives a more accurate energy spectrum, for high
levels. For $\beta=1\cdot10^{-4}$ (the value used in Fig. \ref{fig:<x(t)>})
the WKB method gives a more accurate result for $n>4$.

\subsection{Explicit calculation of $\left\langle \widehat{x}\left(t\right)\right\rangle $
and $\left\langle \hat{p}\left(t\right)\right\rangle $\label{sub:Explicit-calculation-of-<x(t)>}}

The initial wavefunction is the ground state of the harmonic oscillator
$\left|n=0\right\rangle $, displaced by $d$ at time $t=0.$ The
displacement operator is $T\left(d\right)=e^{-i\hat{p}d}$. The displacement
of the harmonic ground state satisfies \cite{Mark_Revivals} 

\begin{equation}
\left\langle m\right|T\left(d\right)\left|0\right\rangle =e^{-\frac{\gamma^{2}}{2}}\frac{\gamma^{m}}{\sqrt{m!}}=e^{-\frac{\gamma^{2}}{2}}C_{m}\left(\gamma\right),\label{eq:Displacement_operator}
\end{equation}

\noindent where 

\begin{equation}
\gamma=\frac{d}{\sqrt{2}}\label{eq:gamma}
\end{equation}

\noindent and

\begin{equation}
C_{n}\left(\gamma\right)=\frac{\gamma^{n}}{\sqrt{n!}}.\label{eq:C_n}
\end{equation}

The displaced state is in fact a coherent state of the harmonic oscillator
Hamiltonian \cite{QM_Messiah} . We assume that the correction to
the harmonic Hamiltonian does not change the eigenstates significantly
and therefore use this basis of eigenstates in future calculations. 

The wavefunction evolution is given by 

\begin{equation}
\left|\psi\left(t\right)\right\rangle =e^{-\frac{\gamma^{2}}{2}}\sum_{n}C_{n}\left(\gamma\right)e^{-iE_{n}t}\left|n\right\rangle 
\end{equation}

\noindent and expectation values of the position and momentum operators
are

\begin{eqnarray}
\left\langle \widehat{x}\left(t\right)\right\rangle  & = & \left\langle \psi\left(t\right)\right|\hat{x}\left|\psi\left(t\right)\right\rangle \label{eq:x_t_beginning}\\
 & = & \sqrt{2}\exp\left(-\gamma^{2}\right)\sum_{n=0}^{\infty}\left[C_{n}\left(\gamma\right)C_{n+1}\left(\gamma\right)\sqrt{n+1}\cos\left(\left(E_{n+1}-E_{n}\right)t\right)\right]\nonumber 
\end{eqnarray}

\noindent and

\begin{equation}
\left\langle \hat{p}\left(t\right)\right\rangle =-\sqrt{2}\exp\left(-\gamma^{2}\right)\sum_{n}\left[C_{n}\left(\gamma\right)C_{n+1}\left(\gamma\right)\sqrt{n+1}\sin\left(\left(E_{n+1}-E_{n}\right)t\right)\right].\label{eq:<p(t)> Gamma}
\end{equation}

The occupation probabilities $|\left\langle n\right|\left.\psi\right\rangle |^{2}$
satisfy a Poisson distribution whose mean and variance are

\begin{equation}
\bar{n}=\overline{\left(n-\bar{n}\right)^{2}}=\gamma^{2},\label{eq:n_hat}
\end{equation}

\noindent where $\gamma$ satisfies (\ref{eq:gamma}). These can be
controlled by the initial displacement. 

\noindent Hence, the weight of (\ref{eq:x_t_beginning}) is concentrated
around $\bar{n}.$ Using the expression (\ref{eq:C_n}) for $C_{n}$
one finds

\begin{equation}
\left\langle \hat{x}\left(t\right)\right\rangle =\sqrt{2}e^{-\gamma^{2}}\gamma\sum_{n}\left(\frac{\gamma^{2n}}{n!}\right)\cos\left(\left(E_{n+1}-E_{n}\right)t\right)
\end{equation}

\noindent and

\begin{equation}
\left\langle \hat{p}\left(t\right)\right\rangle =-\sqrt{2}e^{-\gamma^{2}}\gamma\sum_{n}\left(\frac{\gamma^{2n}}{n!}\right)\sin\left(\left(E_{n+1}-E_{n}\right)t\right).
\end{equation}

\noindent Expanding to the second order in the deviation from $\bar{n}$
and using the Stirling formula \cite{Stirling_book} results in 

\begin{eqnarray}
\left\langle \widehat{x}\left(t\right)\right\rangle  & = & \frac{1}{\sqrt{\pi}}\sum_{n}e^{-\frac{1}{2\gamma^{2}}\left(n-\overline{n}\right)^{2}}\cos\left(\left(E_{n+1}-E_{n}\right)t\right)\label{eq:x(t) after Stirling}\\
 & = & \frac{1}{\sqrt{\pi}}\Re\left(\sum_{n}e^{-\frac{1}{2\gamma^{2}}\left(n-\overline{n}\right)^{2}}e^{-i\left(E_{n+1}-E_{n}\right)t}\right)\nonumber 
\end{eqnarray}

\noindent and

\begin{eqnarray}
\left\langle \hat{p}\left(t\right)\right\rangle  & = & -\frac{1}{\sqrt{\pi}}\sum_{n}e^{-\frac{1}{2\gamma^{2}}\left(n-\bar{n}\right)^{2}}\sin\left(\left(E_{n+1}-E_{n}\right)t\right)\label{eq:p(t) after Stirling}\\
 & = & \frac{1}{\sqrt{\pi}}\Im\left(\sum_{n}e^{-\frac{1}{2\gamma^{2}}\left(n-\overline{n}\right)^{2}}e^{-i\left(E_{n+1}-E_{n}\right)t}\right),\nonumber 
\end{eqnarray}

\noindent respectively.

\noindent From (\ref{eq:Energy_spectrum_full}) we find 

\begin{eqnarray}
\Delta E_{n}=E_{n+1}-E_{n} & = & 1+\frac{3\beta}{4}\left(n+1\right)-\beta^{2}\left(\frac{51}{32}n+\frac{51}{64}n^{2}+\frac{119}{256}\right).\label{eq:Freq_difference}
\end{eqnarray}

\noindent In (\ref{eq:x(t) after Stirling}) and (\ref{eq:Freq_difference})
it is useful to change to the variable 
\begin{equation}
n^{\prime}=n-\bar{n}.\label{eq:n_tag}
\end{equation}
In terms of this variable

\begin{eqnarray}
\Delta E_{n} & = & b_{0}+b_{1}n^{\prime}+b_{2}n^{\prime2},\label{eq:delta_En_1}
\end{eqnarray}

\noindent where

\begin{eqnarray}
b_{0} & = & E_{\bar{n}+1}-E_{\bar{n}}\label{eq:b_zero}\\
 & = & 1+\frac{3}{4}\beta\left(\overline{n}+1\right)-\beta^{2}\left(\frac{119}{256}+\frac{51}{32}\overline{n}+\frac{51}{64}\overline{n}^{2}\right),\nonumber 
\end{eqnarray}

\begin{equation}
b_{1}=\frac{3}{32}\left(8\beta-17\beta^{2}-17\overline{n}\beta^{2}\right),\label{eq:b_1}
\end{equation}

\noindent and 

\begin{equation}
b_{2}=-\frac{51}{64}\beta^{2}.\label{eq:b_2}
\end{equation}

\noindent It is useful to define the sum

\begin{eqnarray}
\widetilde{S}^{x} & = & \sum_{n^{'}=-\bar{n}}^{\infty}e^{-\frac{n^{\prime2}}{2\gamma^{2}}}e^{it\left(b_{1}n^{\prime}+b_{2}n^{\prime2}\right)}\label{eq:S_x_sum_part}
\end{eqnarray}

\noindent In this notation,

\begin{equation}
\left\langle \hat{x}\left(t\right)\right\rangle =\frac{1}{\sqrt{\pi}}\Re\left(e^{-ib_{0}t}\widetilde{S}^{x}\right)
\end{equation}

\noindent and

\begin{equation}
\left\langle \hat{p}\left(t\right)\right\rangle =\frac{1}{\sqrt{\pi}}\Im\left(e^{-ib_{0}t}\widetilde{S}^{x}\right).
\end{equation}

\noindent In the leading order in $\beta$, we see that the sum (\ref{eq:S_x_sum_part})
exhibits revivals at integer multiples of 

\begin{equation}
T_{r}=\frac{2\pi}{b_{1}}.\label{eq:Revival_time}
\end{equation}

\noindent Using (\ref{eq:b_zero}), these are superimposed on the
rapid oscillations of period

\begin{equation}
T_{osc}=\frac{2\pi}{b_{0}}.\label{eq:T_osc}
\end{equation}

\noindent This period is approximately corresponding to the classical
harmonic oscillator period (equals to $2\pi$ in our dimensionless
units). 

\noindent In the leading order in $\beta$ (using (\ref{eq:Revival_time})
and (\ref{eq:b_1})) 

\begin{equation}
T_{r}=\frac{8\pi}{3\beta}.\label{eq:revival_short}
\end{equation}

\noindent We set $t=mT_{r}+\tau$ around the $m^{th}$ revival, where
$-\frac{1}{2}T_{r}<\tau<\frac{1}{2}T_{r}$ .Hence, the sum (\ref{eq:S_x_sum_part})
can be written in the form 
\begin{equation}
\widetilde{S}^{x}=\sum_{m}\widetilde{S}_{m}^{x}\label{eq:Sum_Sm}
\end{equation}
 with

\begin{equation}
\widetilde{S}_{m}^{x}=\sum_{n^{'}=-\bar{n}}^{\infty}e^{-\frac{n^{\prime2}}{2\gamma^{2}}}e^{i\left(b_{1}\cdot n^{\prime}+b_{2}\cdot n^{\prime2}\right)\left(mT_{r}+\tau\right).}\label{eq:S_m}
\end{equation}

\noindent $\widetilde{S}_{m}^{x}$ can be approximated by the integral 

\begin{equation}
\widetilde{S}_{m}^{x}=\int_{-\infty}^{\infty}e^{-\frac{n^{\prime2}}{2\gamma^{2}}}e^{i\left(b_{1}\cdot n^{\prime}\tau+b_{2}\cdot n^{\prime2}\left(mT_{r}+\tau\right)\right)}dn^{\prime}.\label{eq:S_x_integral}
\end{equation}

The reason is that only the terms with small $n^{\prime}$ contribute
substantially to the sum and the differences between adjacent terms
in the sum are small, for $\tau\ll T_{r}$. However, when $T_{c}/T_{r}$
is small, the revival is within these time limits, and outside the
fuction equals to zero. The integral (\ref{eq:S_x_integral}) is a
Gaussian integral calculated using the methods implemented in \cite{Hagar_Collapse_Revival}.

\noindent The result to the leading order in $\beta$ is, using (\ref{eq:x(t) after Stirling}),
(\ref{eq:p(t) after Stirling}), (\ref{eq:b_zero}) and the integrals
(\ref{eq:S_x_integral})

\begin{eqnarray}
\left\langle \hat{x}\left(t\right)\right\rangle  & = & f_{env}\left(t\right)\cos\left(\left(1+\frac{3}{4}\beta\left(\bar{n}+1\right)\right)t\right)\label{eq:x(t) final}
\end{eqnarray}

\noindent and

\begin{equation}
\left\langle \hat{p}\left(t\right)\right\rangle =-f_{env}\left(t\right)\sin\left(\left(1+\frac{3}{4}\beta\left(\bar{n}+1\right)\right)t\right),\label{eq:p(t) final}
\end{equation}

\noindent where

\begin{equation}
f_{env}\left(t\right)=\sqrt{2}\gamma\sum_{m}e^{-\frac{1}{2}\left(\frac{t-mTr}{\sigma}\right)^{2}}\label{eq:env_f}
\end{equation}

\noindent and

\begin{equation}
\frac{1}{\sigma}=\gamma b_{1}\label{eq:1_over_sigma}
\end{equation}

\noindent and $\gamma$ is given by (\ref{eq:gamma}). 

\noindent In the leading order in $\beta$ the revival time $T_{r}$
is given by (\ref{eq:revival_short}). In this order the width $\sigma$
satisfies 

\begin{equation}
\frac{1}{\sigma}=\frac{3}{4}\gamma\beta\label{eq:one_over_sigma}
\end{equation}

\noindent The collapse time is defined as the time when the envelope
$f_{env}\left(t\right)$ reaches $1/e$ of its initial value. Therefore
it takes the value 

\begin{eqnarray}
T_{c} & = & \sqrt{2}\sigma\\
 & = & \frac{T_{r}}{\sqrt{2}\pi\gamma}\nonumber 
\end{eqnarray}

\noindent where we used (\ref{eq:1_over_sigma}) and (\ref{eq:Revival_time})
that are correct to the leading order in $\beta$ and the next one. 

\noindent Using (\ref{eq:gamma}) we find 

\begin{eqnarray}
T_{c} & = & \frac{T_{r}}{\pi d}.\label{eq:collapse_time}
\end{eqnarray}

The dynamics of $\left\langle \hat{x}\left(t\right)\right\rangle $
and $\left\langle \hat{p}\left(t\right)\right\rangle $ exhibit two
different time scales. It displays rapid oscillations with period
of $T_{osc}$ (\ref{eq:T_osc}) superimposed by a slowly varying envelope
$f_{env}\left(t\right)$. This envelope is a sum of Gaussian functions,
separated by time intervals of $T_{r}$ (\ref{eq:Revival_time}).
The collapse time of each Gaussian is $T_{c}$ of (\ref{eq:collapse_time}). 

\begin{figure}[H]
\subfloat[]{\centering{}\includegraphics[clip,scale=0.36]{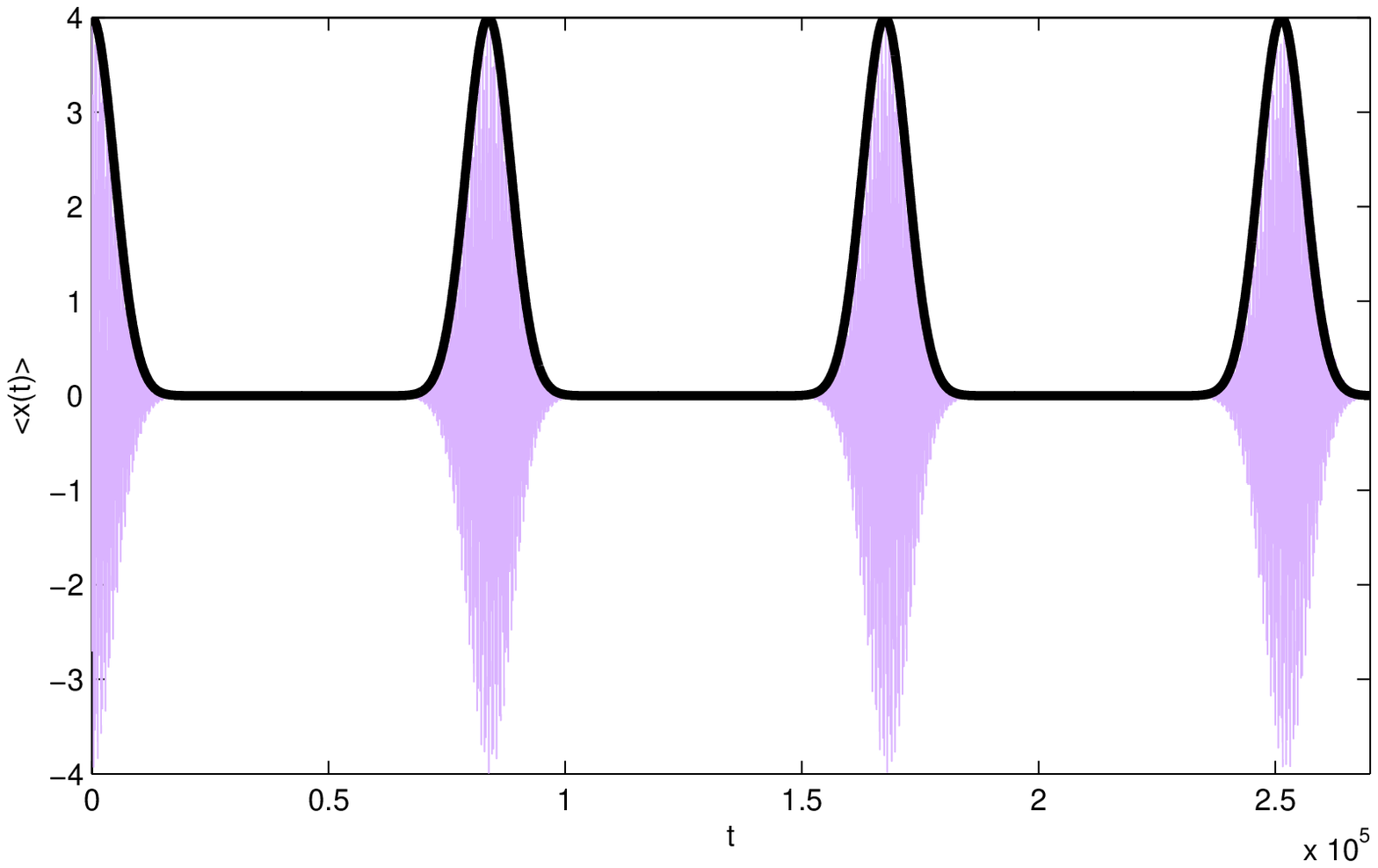}}\subfloat[]{\centering{}\includegraphics[scale=0.32]{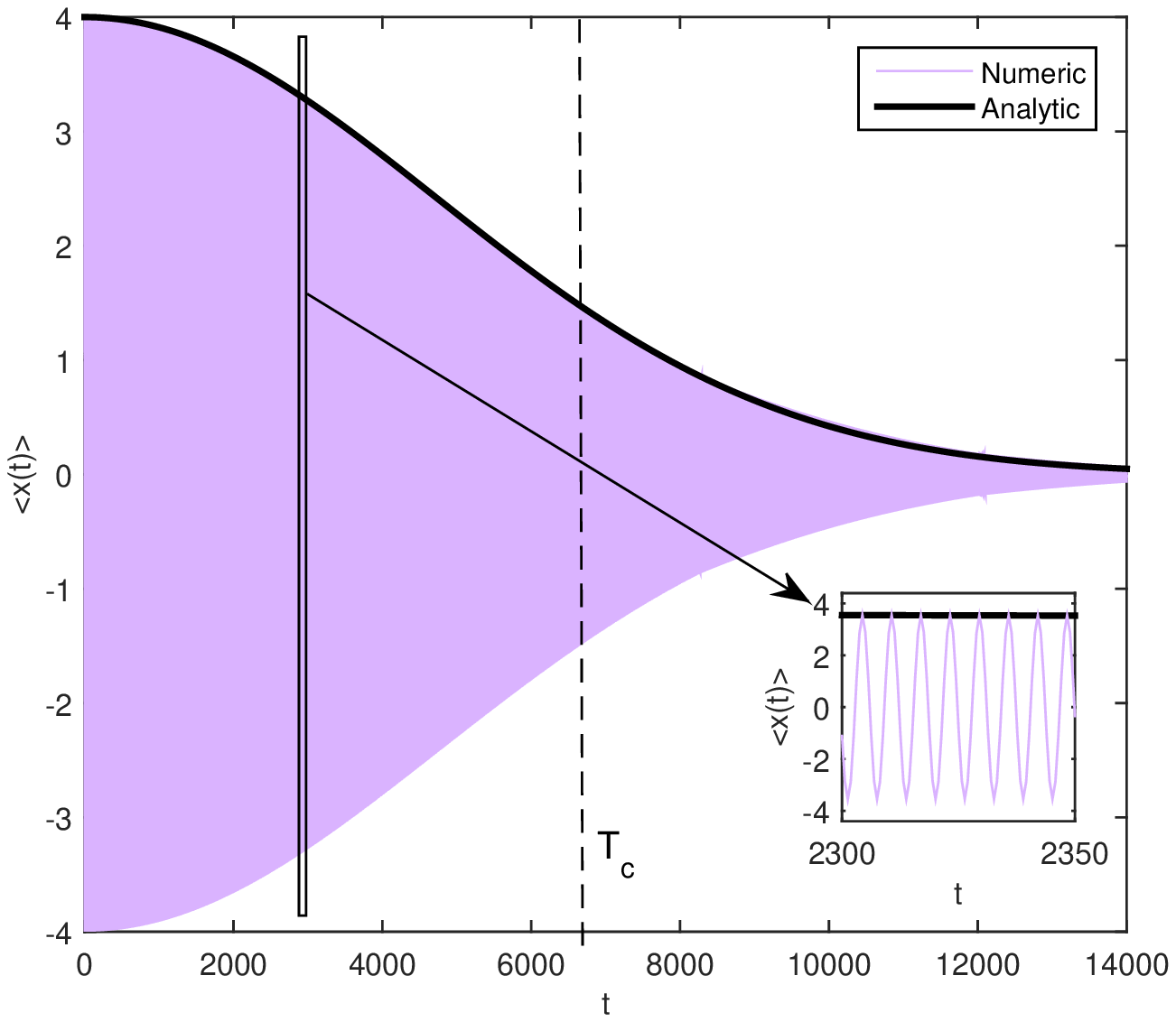}}\hfill{}\protect\caption{\label{fig:<x(t)>}Numerical results based on the solution of the
Schrödinger equation with the Hamiltonian (\ref{eq:Rescaled_Hamiltonian})
(thin purple) and the envelope function $f_{env}$ of (\ref{eq:env_f})
(solid black) for $\left\langle \hat{x}\left(t\right)\right\rangle $
with $\beta=1\cdot10^{-4}$ and $d=4$: (a) three first revivals$\hspace{0.5mm}$(b)
zoom-in on the collapse and the fast oscillations}
\end{figure}

Fig. \ref{fig:<x(t)>} shows the collapse and revival of position,
calculated numerically and its envelope calculated analytically using
(\ref{eq:env_f}), $T_{r}$ of (\ref{eq:revival_short}) and $1/\sigma$
of (\ref{eq:one_over_sigma}) for $d=4$ and $\beta=1\cdot10^{-4}.$
We can see good agreement between the numerical and analytical results. 

For this small value of $\beta$ the leading order in $\beta$ is
sufficient to reproduce the exact numerical results to a high degree
of accuracy. However, for larger values of $\beta$ relevant for the
experimental realizations presented in Sec. \ref{sec:Experimental}
it is instructive to add the next order correction in $\beta.$ The
envelope function takes the form (\ref{eq:env_f}) but with 

\begin{eqnarray}
T_{r} & \simeq & \frac{8\pi}{3\beta\left(1-\frac{17}{8}\beta-\frac{17}{8}\gamma^{2}\beta\right)}\label{eq:Tr_full}\\
 & \simeq & \frac{8\pi}{3\beta}\left(1+\frac{17}{8}\beta+\frac{17}{8}\gamma^{2}\beta\right)\nonumber 
\end{eqnarray}

\noindent and

\begin{equation}
\frac{1}{\sigma}=\frac{3}{4}\gamma\beta\left(1-\frac{17}{8}\beta-\frac{17}{8}\gamma^{2}\beta\right),\label{eq:one_over_sigma_full}
\end{equation}

\noindent where (\ref{eq:b_1}), (\ref{eq:gamma}), (\ref{eq:Revival_time})
and (\ref{eq:1_over_sigma}) were used. 

\noindent There are also corections of order $\beta^{2}$ to the phases
in the sine and cosine functions in (\ref{eq:x(t) final}) and (\ref{eq:p(t) final}),
respectively. These are of no special interest. 

\section{Possible experimental realization for cold atoms\label{sec:Experimental}}

In order to observe impressive collapse and revival structures it
is required that the collapse time $T_{c}$ is much shorter than the
revival time $T_{r},$ and that the revival time is much shorter than
the coherence time, namely, the time over which the system is phase
coherent.

\subsection{Crossed beam trap}

In the previous section we assumed the trap to be one dimensional
and nearly harmonic, namely, the anharmonic correction is assumed
to be small. In order to consider a three dimensional trap effectively
one dimensional, it is required that only the ground state of the
transverse motion is populated. 

Let the direction of the trap and the transverse direction be $z$
and $x,$ respectively. In order to satisfy the quasi-1D condition,
the maximal occupied energy in the $z$ direction of the trap must
satisfy

\begin{equation}
n\ll n_{max}
\end{equation}

\noindent where 

\begin{eqnarray}
n_{max} & = & \frac{\omega_{x}}{\omega_{z}}.
\end{eqnarray}

\noindent \begin{flushleft}
$\omega_{z}$ and $\omega_{x}$ are the frequencies of the longitudinal
and transverse directions, respectively. 
\par\end{flushleft}

In order to have many available energies in the $z$ direction while
the system is still considered one dimensional it is required $\omega_{x}/\omega_{z}\gg1.$

\begin{flushleft}
A trap of this type can be realized using perpendicular crossed Gaussian
beams resulting in a dipolar trap \cite{Traps_Muller}, where the
strongly confining beam is along the $x$ axis and the other one is
along the $z$ axis. 
\par\end{flushleft}

We turn now to estimate the various quantities for Rubidium. The wavelength
of the laser and atomic mass of Rubidium are $\lambda=1064\mathrm{nm}$
and $m_{Rb}=1.4432\cdot10^{-25}\mathrm{kg},$ respectively. Using
cross beam lasers with waists $r_{\parallel}=18\lambda$, $r_{\perp}=6\lambda$
and intensities $P_{\perp}=50\mathrm{W}$ and $P_{\parallel}=5\mathrm{W},$
one finds $\beta=3.47\cdot10^{-5},$ $\omega_{z}=5.36\cdot10^{4}\mathrm{s^{-1}},$
$\omega_{x}=1.04\cdot10^{6}\mathrm{s^{-1}}$ and $n_{max}=19.$ This
leads to a revival time of $T_{r}^{\prime}=4.37\mathrm{s}.$ This
revival time can be achieved experimentally. In fact, there may be
situations where coherence can be maintained for much longer times
\cite{Jeff}. 

If we require that initially, $\bar{n}$ must satisfy

\begin{equation}
\bar{n}+3\sqrt{\left(n-\bar{n}\right)^{2}}<n_{max},
\end{equation}

\noindent \begin{flushleft}
then $\gamma<3.11$ (see (\ref{eq:n_hat})) and $d<4.4.$ 
\par\end{flushleft}

\subsection{Optical lattice potential}

Consider a particle moving in the potential 

\begin{equation}
V\left(x^{\prime}\right)=K^{\prime}\left(1-\cos\left(qx^{\prime}\right)\right)\label{eq:optical_potential}
\end{equation}

\noindent where $q=\frac{2\pi}{\lambda}$. 

Assume the time is sufficiently short and $K^{\prime}$ sufficiently
large so that tunneling can be ignored. Then the dynamics can be considered
as the one taking place in an anharmonic well discussed in Section
\ref{sec:Collapse-and-Revival}. For $qx^{\prime}\ll\frac{\pi}{2}$,
this potential is approximated well by

\begin{equation}
V\left(x^{\prime}\right)\simeq\frac{1}{2}K^{\prime}q^{2}x^{\prime2}-\frac{K^{\prime}}{4!}q^{4}x^{\prime4}.\label{eq:potential_harmonic_trap}
\end{equation}

\noindent Therefore, the harmonic part of the potential can be written
as

\begin{equation}
\frac{1}{2}m^{\prime}\omega_{0}^{\prime2}x^{\prime2}=\frac{1}{2}K^{\prime}q^{2}x^{\prime2},
\end{equation}

\noindent and hence the frequency of the harmonic oscillator is

\begin{equation}
\omega_{0}^{\prime}=\sqrt{\frac{K^{\prime}}{m^{\prime}}}q.\label{eq:omega_0_unscaled}
\end{equation}

\noindent The anharmonic part is of the form

\begin{equation}
\frac{1}{4}\beta^{\prime}x^{\prime4}=-\frac{K^{\prime}}{4!}q^{4}x^{\prime4}
\end{equation}

\noindent Therefore the anharmonicity coefficient is

\begin{eqnarray}
\beta^{\prime} & = & -\frac{K^{\prime}q^{4}}{6}.
\end{eqnarray}

\noindent In dimensionless units, the anharmonicity coefficient becomes,
using (\ref{eq:omega_0_unscaled}) and (\ref{eq:beta_rescaled-1})

\begin{eqnarray}
\beta & = & -\frac{q\hbar}{6m^{\prime\frac{1}{2}}K^{\prime\frac{1}{2}}}.\label{eq:beta_rescaled}
\end{eqnarray}

\noindent At time $t=0$ we shift the ground state of a harmonic oscillator
laterally by $d^{\prime}$ so that 

\begin{equation}
qd^{\prime}=\alpha\pi.\label{eq:alpha}
\end{equation}

The displacement cannot be larger than $\frac{\lambda}{4},$ therefore
from (\ref{eq:alpha}), $\alpha$ should be smaller than $1/2$ and
takes a value so that (\ref{eq:potential_harmonic_trap}) describes
the potential (\ref{eq:optical_potential}) with little deviation. 

\noindent Hence, the dimensionless displacement is

\begin{equation}
d=\alpha\pi\frac{\left(K^{\prime}m^{\prime}\right)^{\frac{1}{4}}}{\left(q\hbar\right)^{\frac{1}{2}}}.\label{eq:d_dimensionless_exp}
\end{equation}

\noindent The various time scales in dimensionless units are the classical
oscillation

\begin{equation}
T_{osc}=2\pi,
\end{equation}

\noindent the revival time (see (\ref{eq:revival_short})) 

\begin{eqnarray}
T_{r} & = & 16\pi\frac{\sqrt{m^{\prime}K^{\prime}}}{q\hbar}
\end{eqnarray}

\noindent and the collapse time (see (\ref{eq:collapse_time})) 

\begin{eqnarray}
T_{c} & = & \frac{16}{\alpha\pi}\frac{\left(K^{\prime}m^{\prime}\right)^{\frac{1}{4}}}{\sqrt{q\hbar}}.
\end{eqnarray}

\noindent In physical units these are

\begin{eqnarray}
T_{osc}^{\prime} & = & \frac{2\pi}{\omega_{0}^{\prime}}=\frac{2\pi}{q}\sqrt{\frac{m^{\prime}}{K^{\prime}},}
\end{eqnarray}

\begin{eqnarray}
T_{r}^{\prime} & = & \frac{T_{r}}{\omega_{0}^{\prime}}=16\pi\frac{m^{\prime}}{q^{2}\hbar}\label{eq:revival_sec}
\end{eqnarray}

\noindent Note that the revival time (\ref{eq:revival_sec}) is independent
of strength of the optical potential. 

\noindent The collapse time is

\begin{eqnarray}
T_{c}^{\prime} & = & \frac{T_{c}}{\omega_{0}^{\prime}}\\
 & = & \frac{16}{\alpha\pi}\left(\frac{m^{\prime3}}{\hbar^{2}K^{\prime}q^{6}}\right)^{\frac{1}{4}}.\nonumber 
\end{eqnarray}

\noindent To see the most pronounced collapse and revival picture,
we would like to increase the ratio

\begin{eqnarray}
\frac{T_{r}}{T_{c}} & = & \pi d\\
 & = & \pi^{2}\alpha\frac{\left(K^{\prime}m^{\prime}\right)^{\frac{1}{4}}}{\left(q\hbar\right)^{\frac{1}{2}}},\nonumber 
\end{eqnarray}

\noindent where we used (\ref{eq:d_dimensionless_exp}). 

It is instructive to calculate the various quantities for some representative
values of the parameters. Following the parameters of \cite{Bloch_Collapse_Revival},
these values are $K^{\prime}=35E_{r},$ $E_{r}=\frac{\hbar^{2}q^{2}}{2m^{\prime}}$
is the recoil energy, $m^{\prime}$ is the mass of a Rubidium atom
and $q=\frac{2\pi}{\lambda}$ where $\lambda=838nm.$ For $K^{\prime}\geq35E_{r}$
tunneling between potential wells can be ignored and these can be
considered isolated \cite{Bloch_Collapse_Revival}. 

One should note that the potential in experiment \cite{Bloch_Collapse_Revival}
is two dimensional. The numbers are presented here to get a general
estimate. We assume here that the motion can be considered decoupled
in the two plane directions. 

For $\alpha=0.25,$ the various quantities take the values $\beta=0.0398,$
$T_{r}=210.3,$ $T_{c}=41.67$, $T_{osc}=6.28,$ $d=1.61$, and in
physical units $\omega_{0}^{\prime}=1.719\cdot10^{5}\frac{1}{sec},$
$T_{r}^{\prime}=1.2msec$, $T_{c}^{\prime}=0.24msec,$ $T_{osc}^{\prime}=36\mu sec,$
$d^{\prime}=0.105\mu m$ and

\begin{eqnarray}
K^{\prime} & = & 7.581\cdot10^{-29}J.\label{eq:K_prime}
\end{eqnarray}

\noindent In experiments, the optical lattice is typically embedded
in a harmonic trap, that is assumed to be modeled by the potential

\begin{equation}
\Delta V=\frac{1}{2}m^{\prime}\omega_{ext}^{\prime2}x^{\prime2},
\end{equation}

\noindent that should be added to the optical potential (\ref{eq:optical_potential}),
where $\omega_{ext}^{\prime}=2\pi\cdot60Hz$ as a typical value \cite{Bloch_Collapse_Revival}
is relatively small compared to the harmonic part of the optical potential
(\ref{eq:optical_potential}) so that $\omega_{ext}^{\prime}\ll\omega_{0}^{\prime}.$
The most important effect of the harmonic trap is the shift of the
minima of the optical lattice potential given by

\begin{equation}
\frac{d}{dx^{\prime}}\left(V+\Delta V\right)|_{x_{min,n}^{\prime}}=0.
\end{equation}

\noindent In the leading order in $\delta_{x},$ the new minima of
the potential are 

\begin{equation}
x_{min,n}^{\prime}=\frac{n\lambda}{2}\left(1-\delta_{x}\right),
\end{equation}

\noindent where

\begin{equation}
\delta_{x}=\frac{\omega_{ext}^{\prime2}}{\omega_{ext}^{\prime2}+\omega_{0}^{\prime2}}.
\end{equation}

\noindent Therefore, the minima are shifted by $\frac{n\lambda}{2}\delta_{x}$
from the original value $\frac{n\lambda}{2}$, without the confinement. 

\noindent Since $\omega_{ext}\ll\omega_{0},$

\begin{equation}
\delta_{x}\simeq\left(\frac{\omega_{ext}^{\prime}}{\omega_{0}^{\prime}}\right)^{2},
\end{equation}

\noindent therefore $\delta_{x}$ is a small parameter. 

\noindent The correction to the optical potential for small distances
from the center of the well is

\begin{equation}
\Delta V\left(\Delta x^{\prime}\right)=\frac{1}{2}m^{\prime}\omega_{0}^{\prime2}x_{min,n}^{\prime2}+\frac{1}{2}m^{\prime}\omega_{ext}^{\prime2}\Delta x^{\prime2}+\frac{1}{6}Kq^{4}\left(\frac{n\lambda}{2}\delta_{x}\right)\Delta x^{\prime3}.
\end{equation}

This is negligible compared to $V\left(\Delta x^{\prime}\right)$
since $\left(\omega_{ext}^{\prime}/\omega_{0}^{\prime}\right)^{2}$
and therefore $\delta_{x}$ are extremely small. Moreover, note that
$\Delta x^{\prime3}$ does not contribute neither to the spectrum
nor to the characteristic time scales in the leading order of perturbation
theory. 

\begin{figure}[H]
\subfloat[]{\centering{}\includegraphics[clip,scale=0.34]{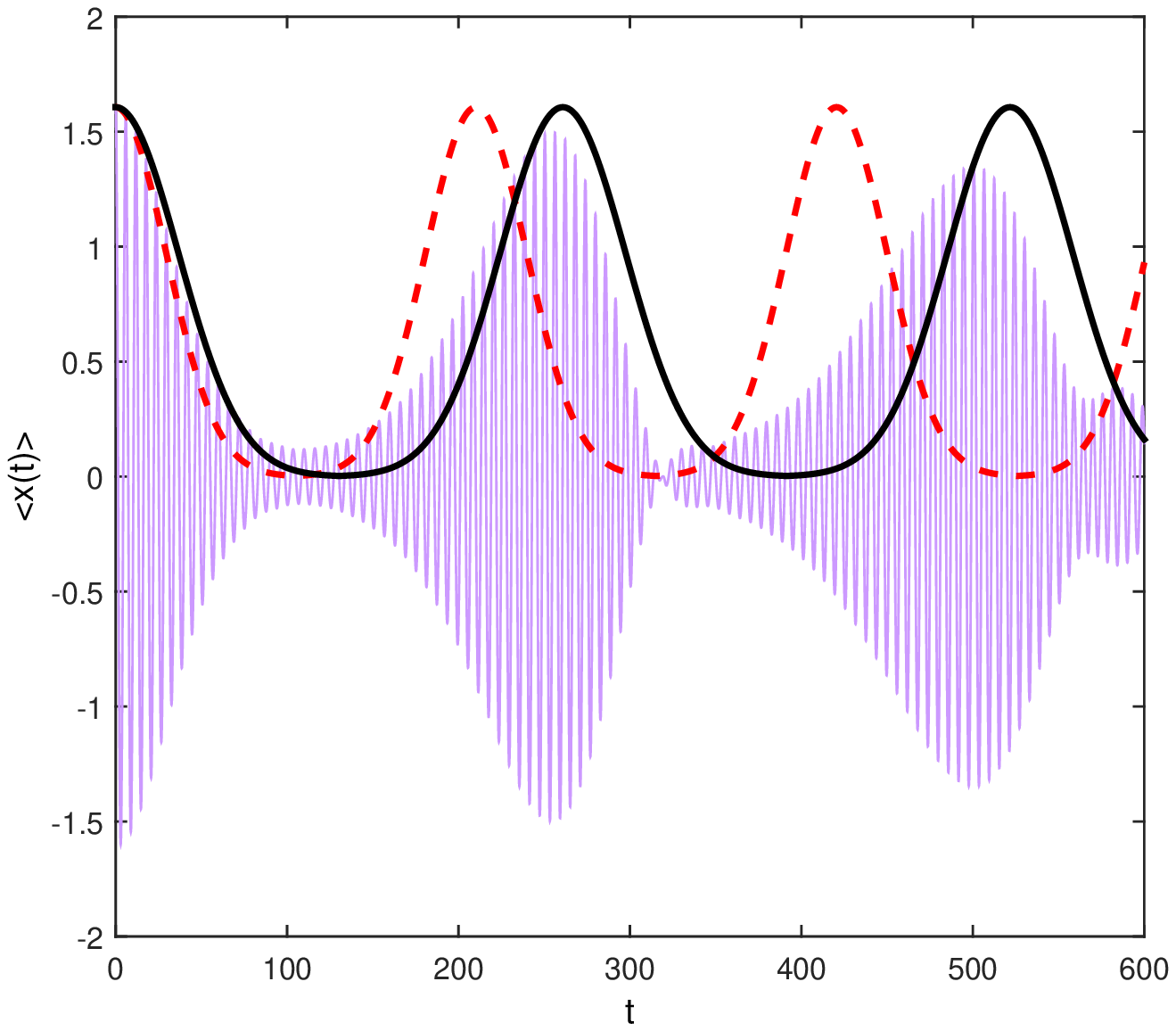}}\subfloat[]{\centering{}\includegraphics[scale=0.34]{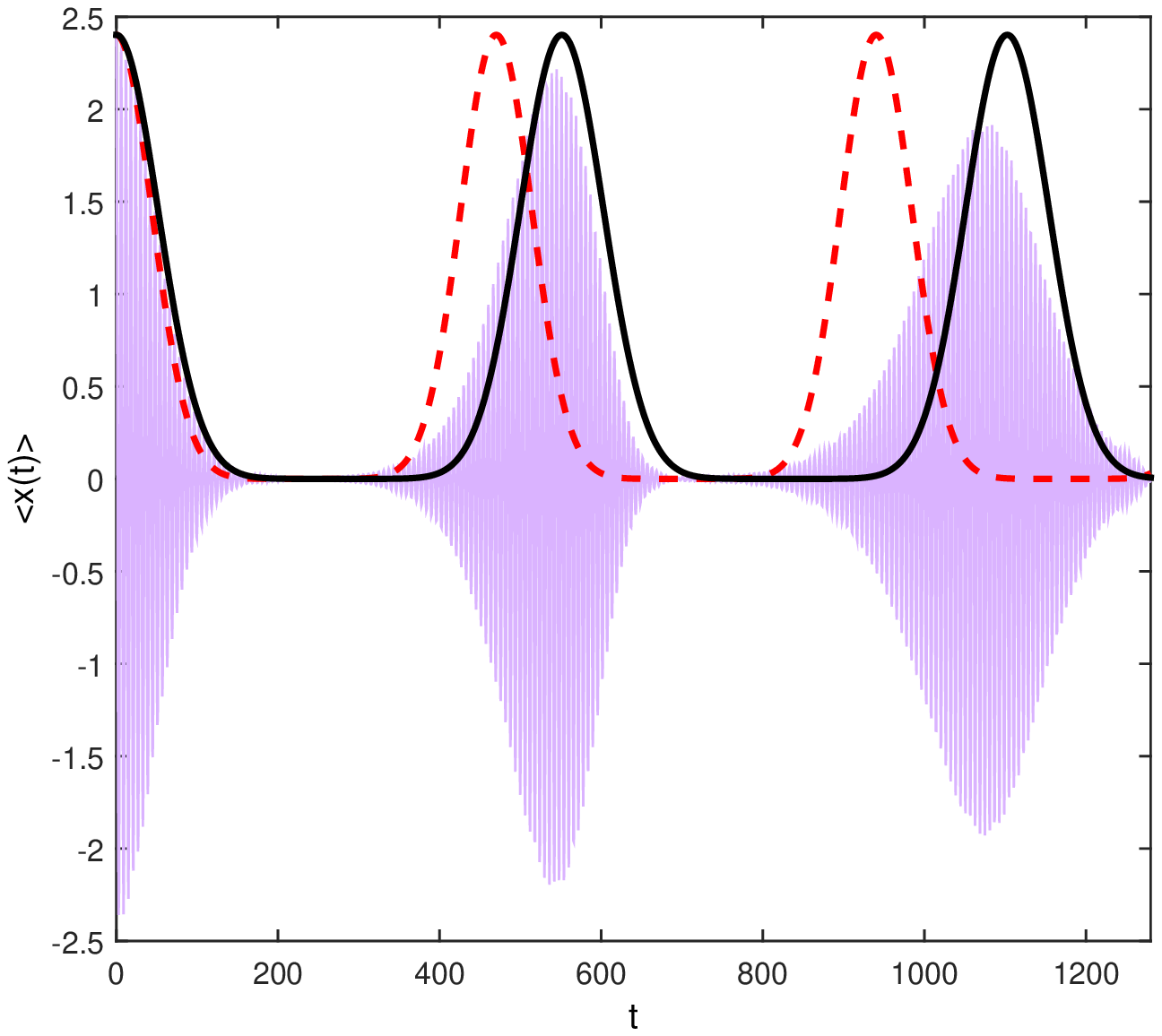}}\hfill{}\subfloat[]{\centering{}\includegraphics[scale=0.34]{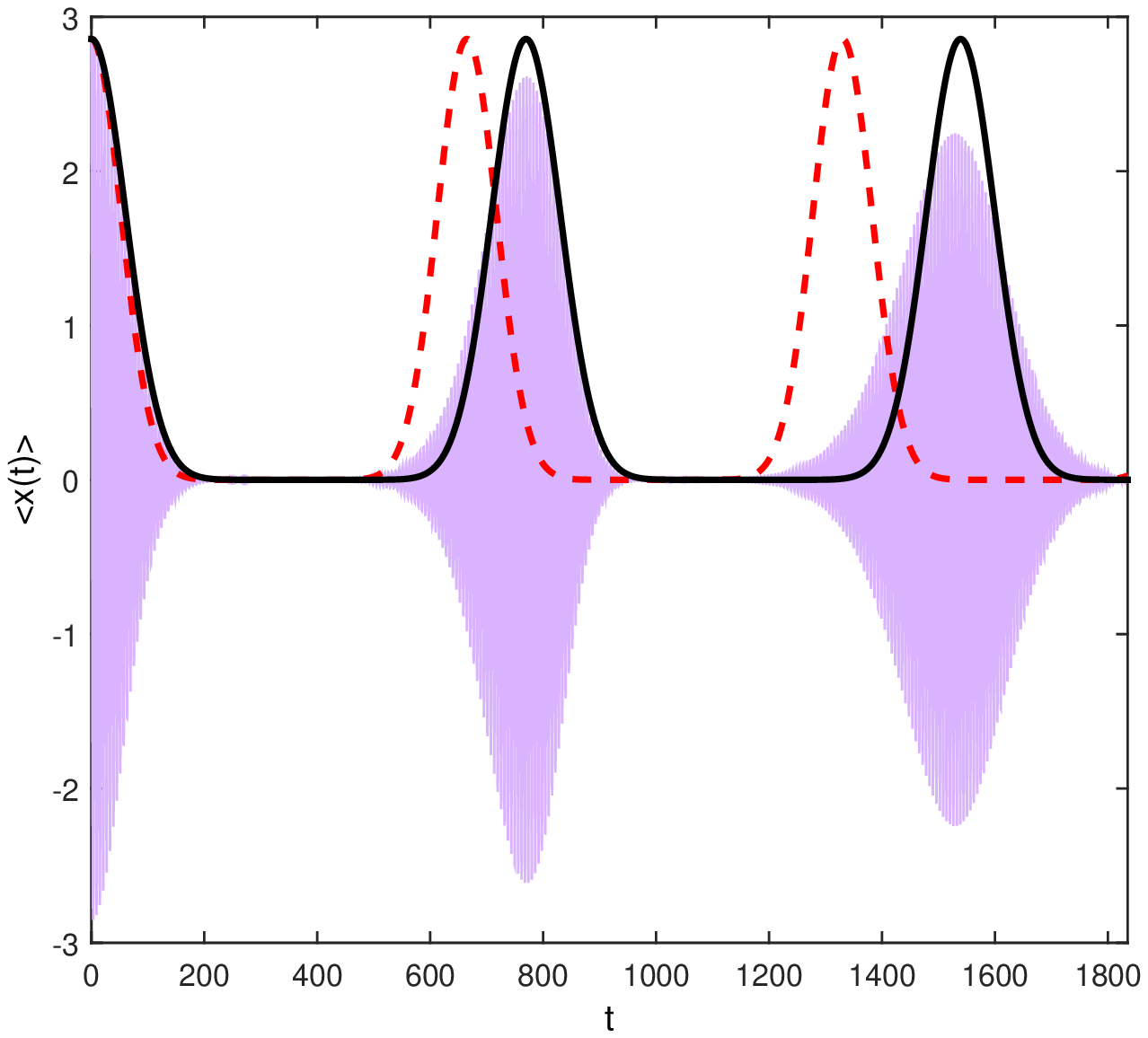}}\protect\caption{\label{fig:Different_K_Collapse_Revival}$\left\langle \hat{x}\left(t\right)\right\rangle $
found numerically as function of time, based on the numerical solution
of the Schrödinger equation with the Hamiltonian (\ref{eq:Rescaled_Hamiltonian})
(thin purple), the analytical envelope predicted by (\ref{eq:env_f})
with $T_{r}$ and $\sigma$ calculated to the first two leading orders
in $\beta$ using (\ref{eq:Tr_full}) and (\ref{eq:one_over_sigma_full})
(solid black) and the envelope predicted by (\ref{eq:env_f}) calculated
to the leading order in $\beta$ using (\ref{eq:revival_short}) and
(\ref{eq:one_over_sigma}) (dashed red) for (a) $K^{\prime}=35E_{r}$,
$\beta=0.0398$ $\hspace{0.5mm}$(b) $K^{\prime}=175E_{r}$, $\beta=0.0178$
$\hspace{0.5mm}$(c) $K^{\prime}=350E_{r}$, $\beta=0.0126$. }
\end{figure}

The collapse and revival picture presented in Sec. \ref{sec:Collapse-and-Revival}
is very impressive. For realistic parameters collapses and revivals
can be observed but the flat regions are much smaller. In Fig. \ref{fig:Different_K_Collapse_Revival}
results corresponding to the ones of Fig. \ref{fig:<x(t)>}a are presented
for the value $K^{\prime}$ of (\ref{eq:K_prime}) as well as for
$K^{\prime}$ increased by a factor of $5$ and $10.$ 

We see that as $\beta$ decreases, the regions where $\left\langle \hat{x}\left(t\right)\right\rangle $
practically vanishes widen. One should note that the width of the
peaks found numerically increases with $m$ due to the blurring effect
which occurs as a result of the last term in (\ref{eq:Freq_difference}).
The correction term of the revival width is of the order of $\beta^{4}$,
hence it could not be estimated in the scope of this work. Therefore,
we expect that a better estimate will be found for higher orders in
$\beta.$ 

The agreement between exact numerical results and approximate analytical
results should improve substantially when the next to the leading
order is included. The difference between the revival times calculated
using (\ref{eq:Tr_full}) and (\ref{eq:revival_short}) is $\Delta T_{r}=\frac{17\pi}{3}\left(1+\bar{n}\right)$,
and $\frac{\Delta T_{r}}{T_{r}}$ , where $T_{r}$ is defined by (\ref{eq:revival_short}),
is proportional to $\beta.$

\section{Summary and Discussion\label{sec:Discussion}}

In this work collapses and revivals were studied for the idealized
model of the anharmonic oscillator presented in Sec. \ref{sec:Collapse-and-Revival}. 

The Hamiltonian 
\begin{equation}
H=\frac{1}{2}p^{2}+\frac{1}{2}x^{2}+\frac{1}{4}\beta x^{4}
\end{equation}
 where $\beta\ll1$, is considered. The initial wavepacket is a ground
state of a harmonic oscillator displaced by $d$ and is concentrated
around the harmonic eigenstate $\bar{n}$ with variance $\bar{n}$
,where $\bar{n}=\gamma^{2}$ and $\gamma=\frac{d}{\sqrt{2}}$ . 

The eigenenergies of this Hamiltonian are expanded around $\bar{n}$,
namely 
\begin{equation}
E_{n}\simeq E_{\bar{n}}+E_{\bar{n}}^{\prime}\left(n-\bar{n}\right)+\frac{1}{2}E_{\bar{n}}^{\prime\prime}\left(n-\bar{n}\right)^{2}+\frac{1}{6}E_{\bar{n}}^{\prime\prime\prime}\left(n-\bar{n}\right)^{3}.
\end{equation}
For the model of this paper this expansion is essentially (\ref{eq:delta_En_1}).
As can be seen from (\ref{eq:b_zero}) - (\ref{eq:b_2}), it is an
expansion in orders of $\beta.$ For small $\beta,$ (\ref{eq:energy_hierarchy})
is satisfied, resulting in a separation of time scales given explicitly
in what follows. Therefore the simple model studied here is representative
of generic systems.

Expectation values of position and momentum are the real and imaginary
parts of 
\begin{equation}
A\left(t\right)=\frac{1}{\sqrt{\pi}}\sum_{n}e^{-\frac{1}{2\gamma^{2}}\left(n-\bar{n}\right)^{2}}e^{-it\left(b_{0}+b_{1}\left(n-\bar{n}\right)+b_{2}\left(n-\bar{n}\right)^{2}\right)},
\end{equation}
where $b_{0},$ $b_{1}$ and $b_{2}$ are defined by (\ref{eq:b_zero}),
(\ref{eq:b_1}) and (\ref{eq:b_2}), respectively. 

The characteristic times are the fast oscillation period, collapse
time and revival time are defined as 
\begin{eqnarray}
T_{osc} & = & \frac{2\pi}{b_{0}},\\
T_{c} & = & \frac{\sqrt{2}}{\gamma b_{1}}
\end{eqnarray}

\noindent and

\begin{equation}
T_{r}=\frac{2\pi}{b_{1}},
\end{equation}

\noindent respectively. 

$A\left(t\right)$ is a rapidly oscillating function superimposed
by an envelope (\ref{eq:env_f}). An approximation of this envelope
in the leading order in $\beta$ gives 
\begin{equation}
f_{env}\left(t\right)=\sqrt{2}\gamma\sum_{m}e^{-\left(\frac{t-mTr}{T_{c}}\right)^{2}},
\end{equation}
 where $m$ is the revival number. 

The simplicity of the model enabled us to derive analytic formulas
(\ref{eq:x(t) final}) and (\ref{eq:p(t) final}) for expectation
values of typical observables. These approximate formulas describe
very well the exact numerical results, as demonstrated in Fig. \ref{fig:<x(t)>}.
In Sec. \ref{sec:Experimental} we demonstrate that the general features
that were found for the idealized models hold also for realistic models.
Fig. \ref{fig:Different_K_Collapse_Revival} then should be compared
to Fig. \ref{fig:<x(t)>}a. There are several systems where one expects
to find collapses and revivals as they share the description (\ref{eq:energy_expansion})
and (\ref{eq:energy_hierarchy}). Each term in (\ref{eq:energy_hierarchy})
is inversely proportional to a time scale of the model, hence the
different time scales are well separated. The results presented in
Sec. \ref{sec:Experimental} show impressive collapse and revival
structure that is extremely close to the one shown in Fig. \ref{fig:<x(t)>}. 

We hope this work will motivate explorations of experimental realizations
that will be as close as possible to the idealized model described
by analytical formulas. We focused on realizations in the field of
atom optics where significant progress was made in recent years and
a high degree of control is possible. An obvious question is : why
to do experiments if a good analytical description is available? In
such situations one can use the analytical solution to explore the
effects of interactions or noise resulting in deviations from the
analytical solution.

In summary, collapses and revivals were found for a large variety
of systems. In the present paper we presented an idealized model that
exhibits this phenomenon, and may enable new directions in its investigation.

\section{Acknowledgements\label{sec:Acknowledgements}}

We thank Immanuel Bloch, Steven Tomsovic, Yoav Sagi and Jeff Steinhauer
for highly informative and stimulating discussions. This work was
supported in part by the Israel Science Foundation (ISF) grants 1028/12
and 931/16, by the US-Israel Binational Science Foundation (BSF) grant
number 2010132 and by the Shlomo Kaplansky academic chair. SF thanks
the Kavli Institute for Theoretical Physics in Santa Barbara for its
hospitality, where this research was supported in part by the US National
Science Foundation (NSF) under grant NSF PHY11-25915. 

\bibliographystyle{h-physrev}
\bibliography{Reference_database2}

\end{document}